\documentclass{article}

\usepackage{PRIMEarxiv}

\usepackage[utf8]{inputenc} % allow utf-8 input
\usepackage[T1]{fontenc}    % use 8-bit T1 fonts
\usepackage[hidelinks]{hyperref}       % hyperlinks
\usepackage{url}            % simple URL typesetting
\usepackage{booktabs}       % professional-quality tables
\usepackage{amsfonts}       % blackboard math symbols
\usepackage{nicefrac}       % compact symbols for 1/2, etc.
\usepackage{microtype}      % microtypography
\usepackage{lipsum}
\usepackage{fancyhdr}       % header
\usepackage{graphicx}       % graphics
\usepackage{float}
\usepackage{multirow}
\usepackage{adjustbox}
\usepackage{seqsplit}
\usepackage{longtable}
\usepackage{enumitem}
\usepackage[version=4]{mhchem}
\title{Enhancing Selection of Climate Tech Startups with AI: A Case Study on Integrating Human and AI Evaluations in the ClimaTech Great Global Innovation Challenge}

\author{
Jennifer Turliuk, 
Alejandro Sevilla, 
Daniela Gorza and
Tod Hynes\\ Massachusetts Institute of Technology
}

\begin{document}
\maketitle

\begin{abstract}
This case study examines the ClimaTech Great Global Innovation Challenge’s innovative approach to selecting climate tech startups by integrating human and AI evaluations. The competition aimed to identify and support the most promising startups in the climate tech sector, and enhance the accuracy and efficiency of the selection process through a hybrid evaluation model. Research has demonstrated that "Data-driven initiatives have been shown to help VC firms reduce gender bias and make better, fairer investment decisions" \cite{noauthor_ddvc_nodate} \cite{noauthor_how_nodate}. Additionally, "Machine learning models have already been proven to outperform human investors in deal screening" \cite{noauthor_ddvc_nodate} \cite{retterath_human_2020}, helping to identify high-potential startups more effectively. By incorporating AI into our evaluation process, we aimed to ensure a more equitable and objective selection of startups. 
The methodology consisted of three phases: an initial review by AI, semi-finals judged by human reviewers, and finals where the weighting was adjusted to prioritize human judgment while still incorporating AI insights. In the first phase, 57 applications were scored by an AI tool built with StackAI and OpenAI’s GPT-4o, selecting the top 36 startups for human review. During the semi-finals, human judges, unaware of AI scores, evaluated the startups based on criteria such as team quality, market potential, and technological innovation. Composite scores were given an equal weight between the number of judges (human and AI) who reviewed a startup (4); this corresponded to a weighting of  75\% (3/4) human scores and  25\% (1/4) AI scores. Meanwhile, with 5 judges at the finals, this resulted in more significant human influence (83.3\%, 5/6) and AI contribution (16.67\%) to ensure a balanced and comprehensive evaluation.
Key findings include a moderate positive correlation between human and AI scores (Spearman’s: 0.47), indicating general alignment with some notable discrepancies. These differences highlight the complementary nature of AI and human judgment, with AI providing additional insights and identifying potential that might not be immediately apparent to human evaluators. An important point to note is that based on the initial ranking of 57 teams, the AI ranked three teams 5.0 and six teams at 4.5 (out of a 5 point scale), and the final 4 teams (primarily selected by the human judges) were within this group. 
The study demonstrates that integrating AI into the evaluation process can streamline and enhance startup assessments, though continuous refinement of AI models is necessary to better align with human intuition and judgment. The hybrid model adopted by the ClimaTech Competition offers a robust framework for future competitions, supporting more impactful and innovative climate solutions by leveraging both human expertise and AI capabilities.

\end{abstract}

% keywords can be removed
%\keywords{Big data, big data analytics, quality frameworks, quality challenges. }

\section{Introduction}

The ClimaTech Great Global Innovation Challenge competition aimed to identify and support the most promising climate tech startups. This study examines how integrating human and AI evaluations helped streamline and enhance the startup selection process and uncover insights. By leveraging AI for initial screening and human expertise for final assessments, the competition sought to enhance the comprehensiveness and efficiency of the startup selection process, and to reduce bias. The process and AI prompts were posted on the competition \href{http://www.climateandenergystartups.org/}{website} in advance of the competition, and there was \href{https://docs.google.com/document/d/13OLVqIbtcfN0Ukv4GvbGhAfFD3uCt6kjfGTMr6tSu6g/edit\#heading=h.v1m19k8ueshj}{a form} available for public feedback.

Research has demonstrated that "Data-driven initiatives have been shown to help VC firms reduce gender bias and make better, fairer investment decisions"  \cite{noauthor_ddvc_nodate} \cite{noauthor_how_nodate}. Additionally, "Machine learning models have already been proven to outperform human investors in deal screening" \cite{noauthor_ddvc_nodate} \cite{retterath_human_2020}, helping to identify high-potential startups more effectively. By incorporating AI into our evaluation process, we aimed to ensure a more equitable and objective selection of startups. The AI screening process also enabled the competition to have an open and global call for applications with a compressed timeline of only \~4 weeks to run the competition. 

The Great Global Innovation Challenge (GGIC) is an innovative initiative designed to directly connect climate-focused startups with potential customers who are actively seeking sustainable solutions as well as potential investors and partners. This competition, developed in collaboration with Climatech.live and supported by the MIT Climate and Energy Prize (MIT CEP), aims not only to identify and accelerate promising startups but also to leverage AI tools to facilitate meaningful connections between these companies and customers, regardless of their competition outcome. The top four teams presented live during the ClimaTech event and received various awards. 

\section{Methodology}

\subsection{Overview:}

As part of Climatech’s Great Global Innovation Challenge, we leveraged advanced artificial intelligence (AI) technology to assess and score each startup on several key dimensions. This process supported a fair, consistent, and comprehensive evaluation of all applicants. The AI utilized natural language processing (NLP) and machine learning models, including OpenAI's GPT-4o, to analyze startup applications based on criteria such as Potential Climate Impact, Solution Creativity, Idea Feasibility, and Team Composition. The scores generated by the AI were used alongside human evaluations to select the startups to present at the challenge.

\subsection{Evaluation Criteria:}

These criteria are based on the selection criteria from the MIT Climate \& Energy Prize, which was founded in 2007 and has had many successful competition alumni.

\begin{itemize}
    \item Potential Climate Impact: The AI assessed the startup's potential environmental benefit in terms of greenhouse gas (GHG) emissions reduction, waste reduction, carbon removal, and the number of lives affected.
    \item Solution Creativity: The AI evaluated the innovativeness and originality of the project, considering whether the idea offers new and creative solutions to existing problems in the field.
    \item Idea Feasibility: The AI analyzed the soundness and realism of the business model, the team’s knowledge of potential obstacles, identification of a specific market or customer, and the scalability of the solution.
    \item Team Composition: The AI assessed the balance and expertise of the team, considering the diversity of backgrounds, technical expertise, and the team’s plan for future growth.
\end{itemize}

In addition, during the final stage of the selection process, there was an AI judge to provide real-time questions for and assessments of the finalist presentations. An AI judge scored the final presentation, and participated in final deliberation and voted alongside human judges. This innovative approach aimed to enhance the objectivity and thoroughness of our selection process.

\begin{figure}
    \centering
    \includegraphics[width=0.5\linewidth]{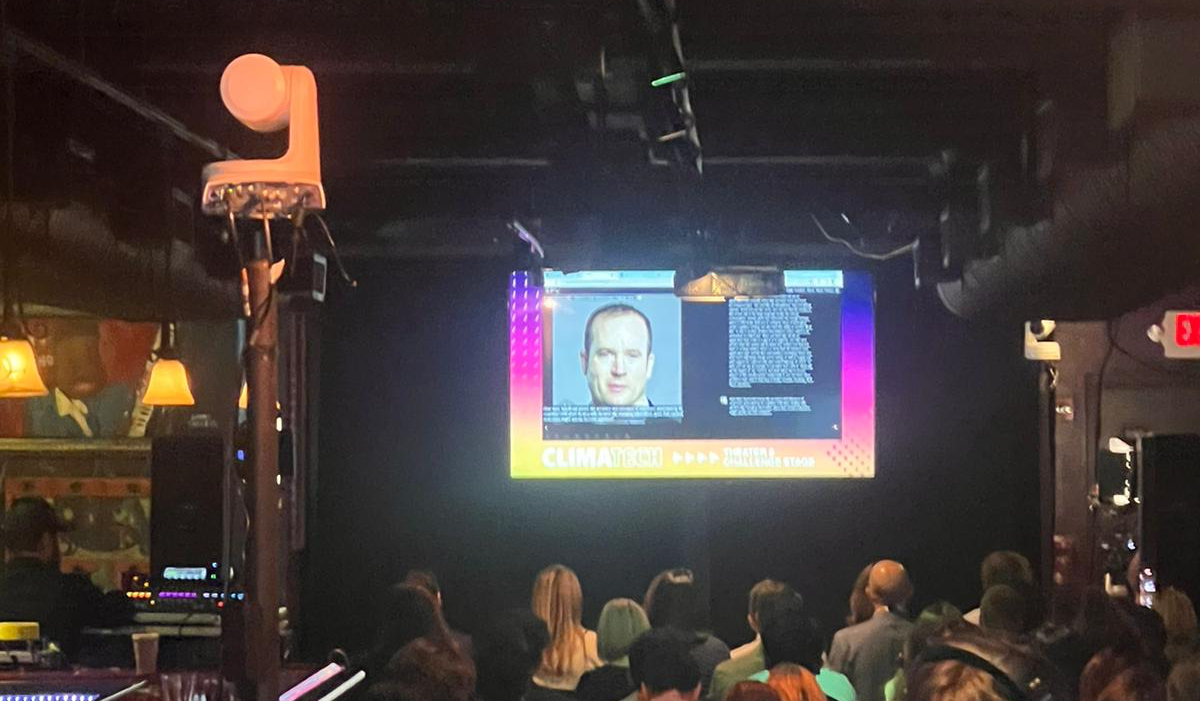}
    \caption{Photo of AI judge}
    \label{fig:1}
\end{figure}

Sample video of AI judge asking a question: \href{https://youtu.be/-5mKKuneOIE}{https://youtu.be/-5mKKuneOIE} (Figure \ref{fig:1}).

\subsection{First Review}

57 applications were reviewed and scored by an AI tool (built with StackAI and OpenAI’s GPT-4o) using prompts based on a specific rubric. The AI provided scores and comments, which were used to select the top 36 startups for human review. This cut-off considered: number of reviewers available, load on reviewers (9 startups/reviewer) in a short time frame, intentional overlap (2-3 reviewers/startup), and thoroughness (wanting to review as many as possible). 

\subsection{Semi-Finals}

The evaluation process combined human expertise with AI assessments. 

Reviewers were randomly assigned to startups and were unaware of AI scores, ensuring unbiased evaluations. This randomization and blinding strengthen the validity of the human review process. Human judges assessed startups using their presentation deck and based on criteria such as team quality, market potential, and technological innovation. AI provided additional scores by analyzing data-driven insights, trends, and historical data. 

The final composite scores were calculated with a 75\% weighting for human scores and 25\% for AI scores.
\[
\textit{Composite Score} = 0.75 \times \textit{Human Score} + 0.25 \times \textit{AI Score}
\]
A second score, a pseudo Z-score, was calculated to account for the variability in the mean and standard deviation for each reviewer (human and AI alike), effectively normalizing and allowing for a fair weighting of each startup. In the ideal scenario, ranking teams by either highest composite scores or composite Z-scores yields the same result; but having different judges for different startups introduces variability that should be compensated for when comparing across startups.

\[
Z\ \textit{score} = \frac{\textit{Composite Score} - \textit{Judge Mean}}{\textit{Judge Standard Deviation}}
\]
\[
\textit{Composite Z Score} = 0.75 \times \textit{Human Z Score} + 0.25 \times \textit{AI Z Score}
\]
To select the 4 finalists, the teams were ranked by the composite score and/or the composite Z-score. The competition’s stipulation required selection of two early-stage and two late-stage startups, with at least one coming from the global south. Hence, the startups who first met that criteria from the ranked list(s) were selected, starting with those with the highest composite score or Z-score.

\subsection{Finals}

In the finals, the weighting was adjusted to 83.33\% for human scores (5 human judges) and 16.67\% for AI scores (1 AI judge) to reflect the relative importance of human judgment in the final selection phase. AI judges were incorporated to provide real-time questions after pitches (based on transcripts) and human judge questions, to provide scores and rank teams based on the presentations and Q\&A sessions, and to participate in judge deliberation. At the beginning of the finals judging session, the 5 human judges determined that in the case of a 3 to 3 tie, the winner would be determined by the 3 human judges, not 2 humans + 1 AI.

\subsubsection{AI Judges Finals Process for ClimaTech’s Great Global Innovation Challenge}

This section reviews the process to incorporate AI Judges in the final judging process. Two AI judges are used to split the persona into one focused on asking questions and providing feedback and a separate persona for actually ranking the teams based on materials and responses. The AI judge will only receive 1 vote even though there are multiple personas.

\subsubsection{AI Judge Integration:}

\begin{enumerate}
\item \textbf{Pitch Evaluation Phase:} 
\begin{itemize}
\item \textbf{AI Judge 1 for Q\&A and Feedback:}

During the pitch, the AI Judge 1 will engage by asking relevant questions to the participants.

It will provide real-time feedback based on the transcript of the presenter during the presentations.

This AI judge will capture and document all the discussion points and responses during the pitch.
\end{itemize}

\item \textbf{Scoring Phase:} 

\begin{itemize}
\item\textbf{AI Judge 2 for Ranking:}
Post-pitch, AI Judge 2 will evaluate the presentation, Q\&A session, and any other relevant materials.

It will score each team on predefined criteria such as Climate Impact, Solution Creativity, Idea Feasibility, and Team Composition. This is the same process for the human judges.

The scores will be based on captured discussions and feedback, ensuring consistency and fairness.

These scores are not shared with the human judges (who are also independently scoring each team with the same process)
\end{itemize}

\item \textbf{Judging Session:}

\begin{itemize}
\item \textbf{Collaborative Decision-Making:}

During the closed-door judging session, AI Judge 1 will provide synthesized insights from the pitch and Q\&A sessions after the human judges have been able to share their thoughts with each other.

It will answer questions like strengths and weaknesses of each team, and comparative analysis of the teams.

All judges will deliberate and discuss, incorporating AI-provided insights to ensure a comprehensive evaluation.
\end{itemize}

\item \textbf{Final Scoring:}

\begin{itemize}
\item \textbf{Consolidated Evaluation:}
At the end of the judging session, the AI will submit its final rankings based on the cumulative data from pitches, feedback sessions, and discussions.
\end{itemize}
\end{enumerate}

These rankings will be used alongside human judges' scores as part of the process to determine the final winners.

\begin{figure}[h]
    \centering
    \includegraphics[width=\linewidth]{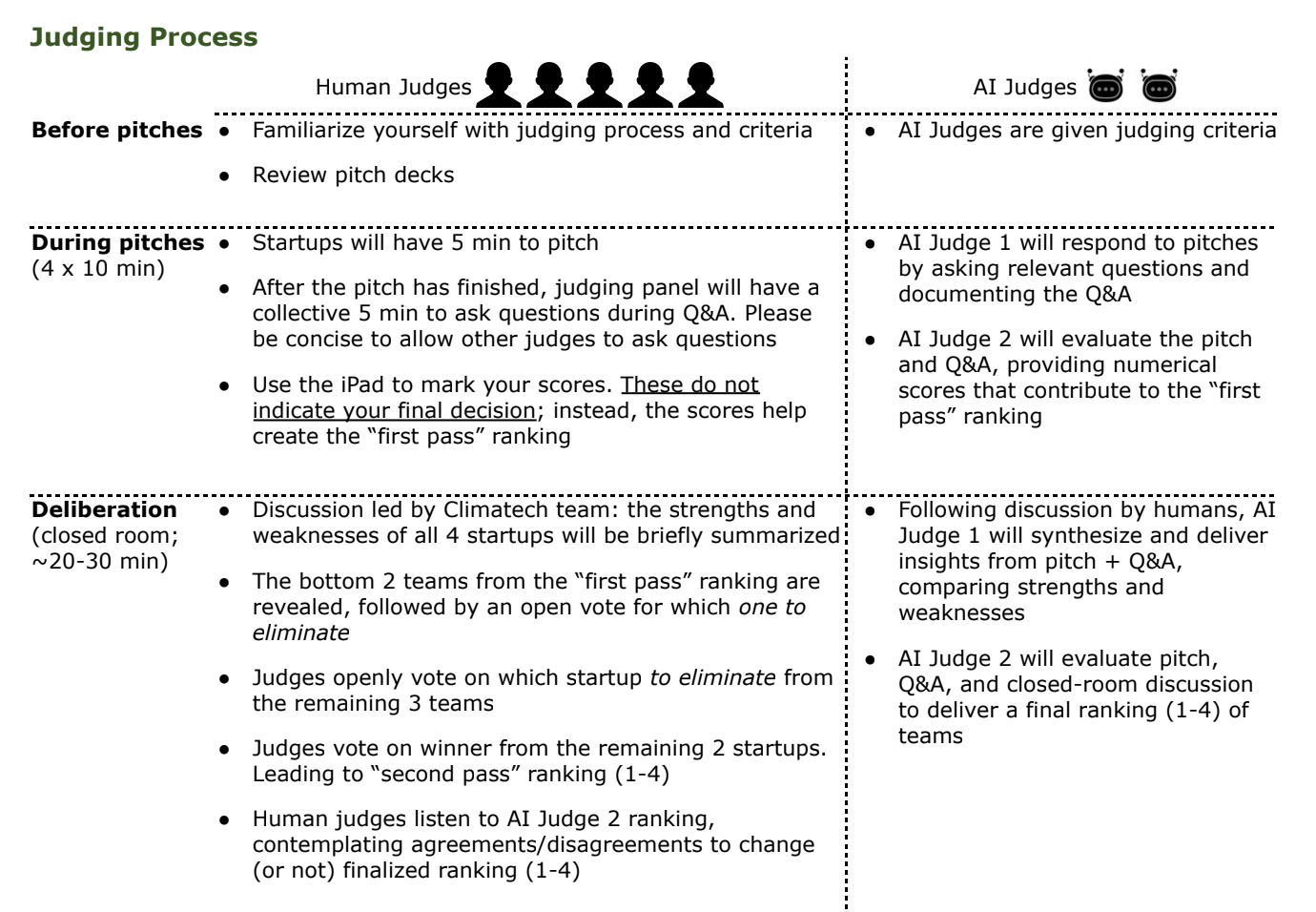}
    \caption{Judging Process}
    \label{fig:2}
\end{figure}

\section{Data Analysis and Findings}

\subsection{Semi-Finals}

\subsubsection{Compiled Results (AI + Humans)}

\begin{figure}[h]
    \centering
    \includegraphics[width=\linewidth]{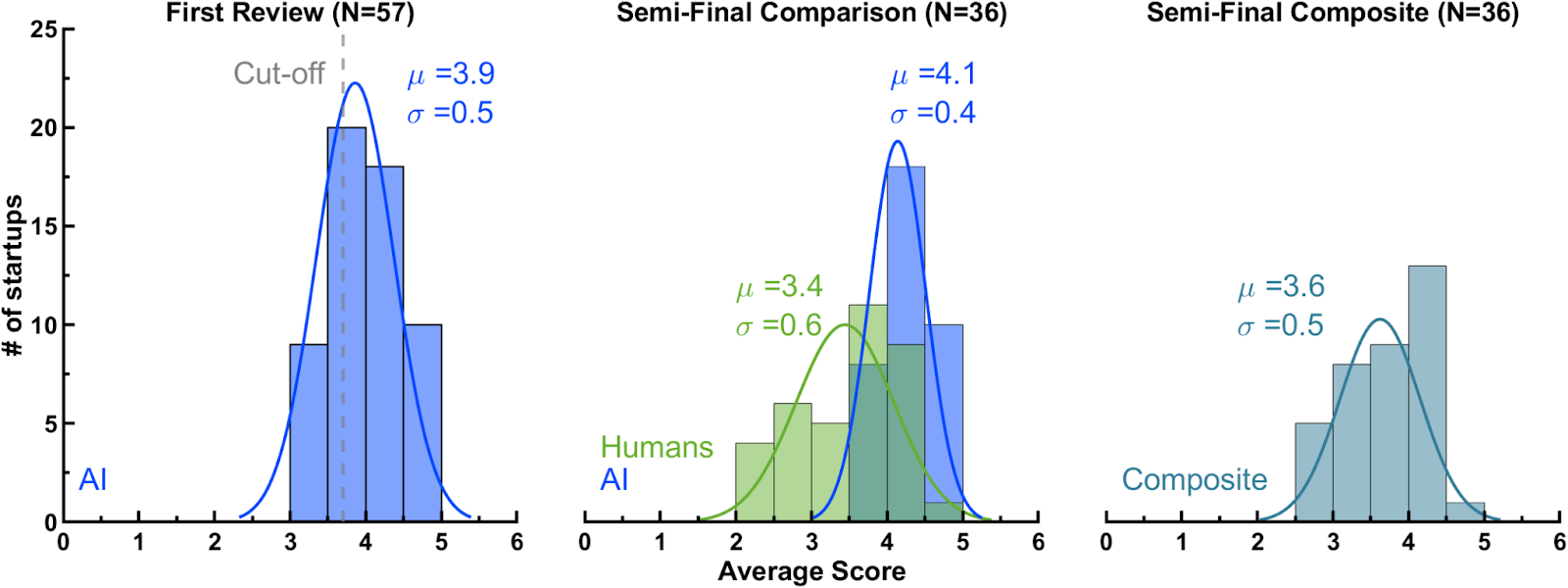}
    \caption{Score Distributions}
    \label{fig:3}
\end{figure}

After selecting the top 36 startups from the AI’s first review of 57 startups, the average score distributions between the human and AI scores can be compared (Figure \ref{fig:3}). Overall, the AI scored these top 36 startups higher than humans (average of 4.1 vs. 3.4), with a tighter distribution of scores (standard deviation of 0.4 vs. 0.6).

The composite scores were derived from a weighted average of human and AI scores (Figure \ref{fig:4}). See Appendix \ref{sec:A} for full results.

\begin{figure}[H]
    \centering
    \includegraphics[width=0.9\linewidth]{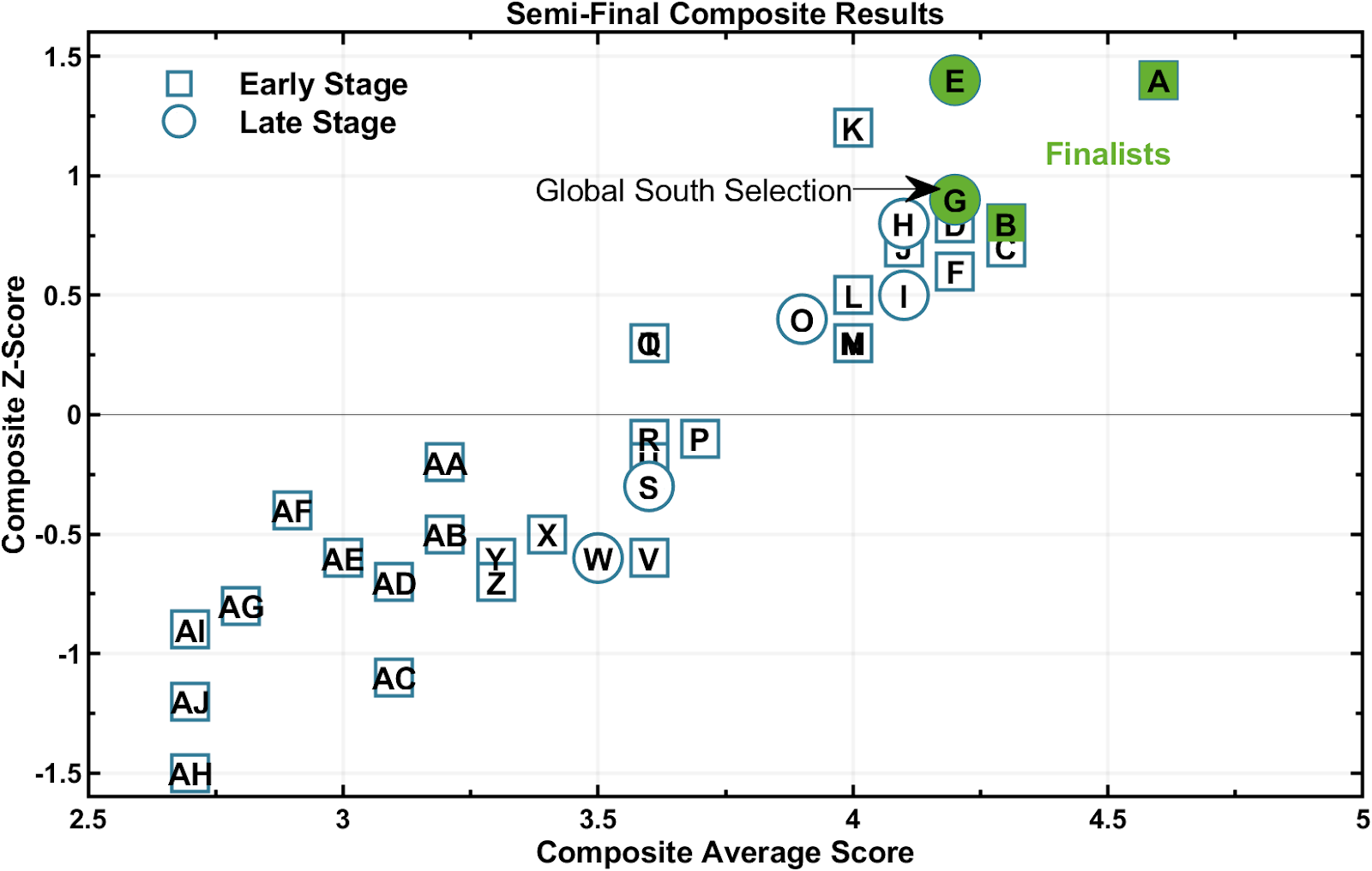}
    \caption{Compiled Results}
    \label{fig:4}
\end{figure}

\subsubsection{Humans vs AI}
\begin{figure}[h]
    \centering
    \includegraphics[width=0.9\linewidth]{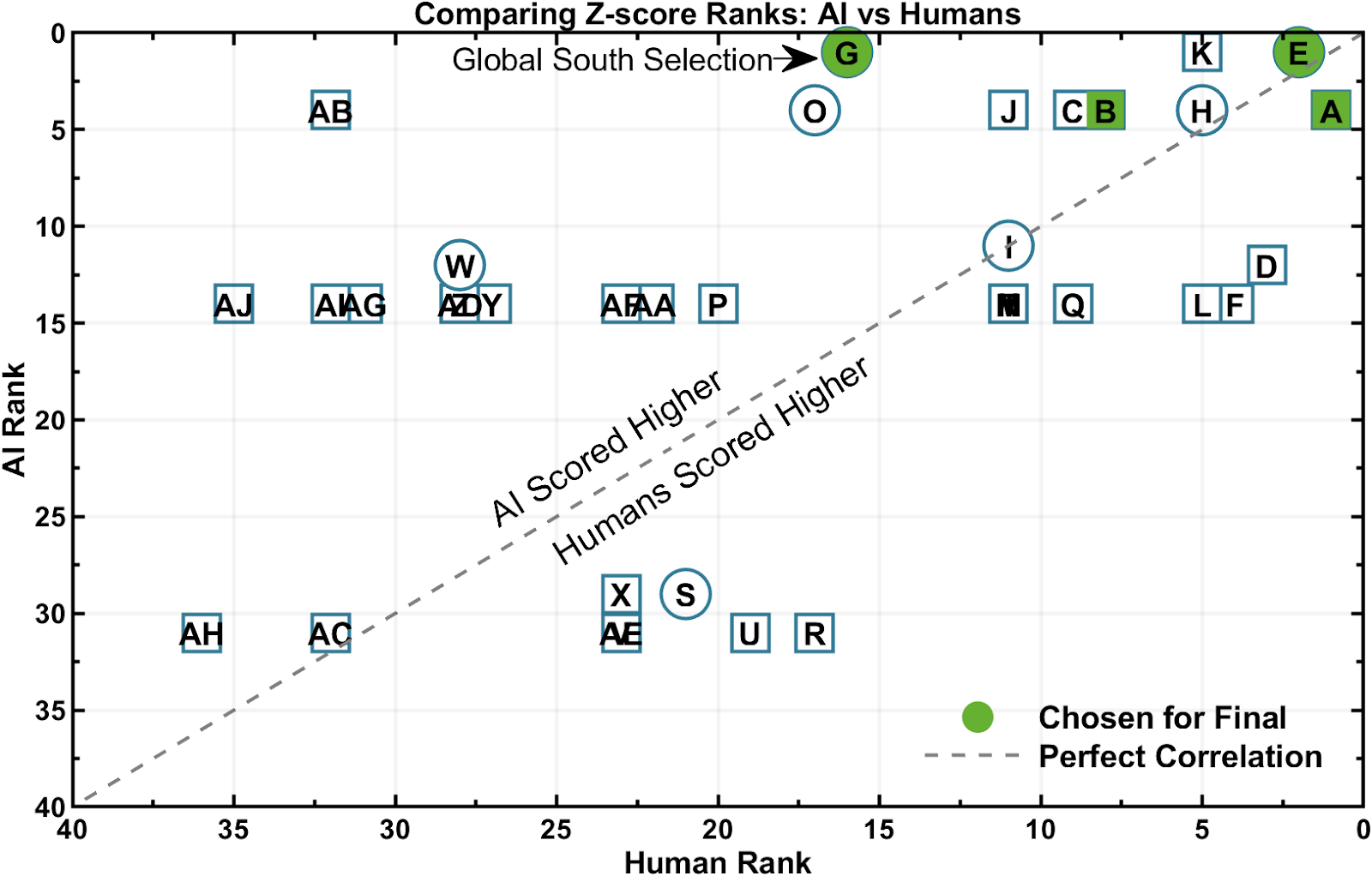}
    \caption{Humans vs AI}
    \label{fig:5}
\end{figure}

The comparison between human and AI Z-score rankings, which accounts for the variability in the averages between judges, provided insights into the correlation and differences between these two evaluation methods (Figure \ref{fig:5}). 
The Spearman’s correlation coefficient of 0.47 indicates a moderate positive relationship between human and AI evaluations, suggesting that while they generally agree, there are notable differences.

\subsection{Finals}

\begin{table}[h]
\centering
\caption{Finalist Team Scores and Rankings}
\label{tab:team-scores}
\resizebox{\linewidth}{!}{%
\begin{tabular}{l  l  l  l  l  l  l }
\toprule
\textbf{Team Name} & \textbf{Human Weighted Score} & \textbf{AI Weighted Score} & \textbf{Human Rank} & \textbf{AI Rank} & \textbf{Composite Score} & \textbf{Composite Rank} \\
\midrule
\textbf{B} & 2.9 & 4.5 & 4 & 2 & 3.2 & 4 \\
\textbf{A} & 3.6 & 4 & 3 & 3 & 3.6 & 3 \\
\textbf{G} & 3.8 & 3.5 & 2 & 4 & 3.8 & 2 \\
\textbf{E} & 3.9 & 4.8 & 1 & 1 & 4.1 & 1 \\
\bottomrule
\end{tabular}%
}
\end{table}

AI Deliberator: Picked startup A over B and picked startup E over G.

Human Deliberation: Picked startup A over B and picked E over G.

\section{Key Insights}

\subsection{Semi-Finals}

\subsubsection{Alignment of Human and AI Scores}

Overall, there was a moderate positive correlation between AI and human rankings (Spearman’s rank correlation coefficient of 0.47), suggesting that AI evaluations are generally in line with human judgments but highlight areas of divergence. This alignment can be leveraged to cross-validate and enhance the reliability of startup evaluations. While the AI provides valuable insights, it should complement rather than replace human judgment. 

Teams A and E both had high human and AI scores, demonstrating strong alignment between human judgment and AI insights. This suggests that when both evaluations align, it indicates a higher confidence in the startup's potential.

There are some noticeable differences between the human and AI scores, such as in Team AB. This team was ranked highly by the AI (top 5), but low by the human judges (bottom 5). Human judges reviewed only the presentation decks, and comments from judges noted limited information was provided on team AB’s presentation deck (\textit{e.g., }no information on team composition). Instead, the AI reviewed text fields of information provided by the startup which may have filled in the gaps from the presentation deck. This example highlights a limitation of this study and suggests room for improvement to ensure consistent information is being reviewed across humans and the AI judge.

\subsubsection{AI as a Complementary Tool:}

AI scores provided additional perspectives, especially in areas where human evaluators might have biases or lack certain insights.

Startups with significant discrepancies between human and AI scores warrant further investigation to understand the reasons behind these differences.

AI scores were also more clustered than human scores, making it difficult to distinguish startups with similar scores using AI scores alone. Further modifications in the starting prompt can help alleviate this clustering. 

\subsubsection{Areas of Discrepancy:}

Some startups received high AI scores but lower human scores and vice versa. This indicates that AI may have identified potential that was not as apparent (or available in the presentation decks) to human judges, suggesting areas for further investigation. These discrepancies highlight the importance of using both AI and human evaluations to capture a comprehensive assessment. 

For instance, Team AB did not provide critical information in the presentation decks that would improve the human review process. Instead, the team only included the key information in the application form reviewed by AI. 

Both AI and human evaluators recognize the importance of team strength and innovative technology. However, AI tends to be more optimistic about feasibility, while human reviewers often point out practical concerns and areas needing clarity. For example, human judges commented they had concerns about the scalability of the solution from Team AJ and scored them low (<2) on idea feasibility, resulting in a low human ranking (bottom 2); instead, the AI scored this team in the top 15 with higher feasibility scores. 

Startups like Team G and Team O showed significant discrepancies between human and AI rankings. These discrepancies highlight areas where human intuition and AI analytics diverge, suggesting the need for further investigation into the evaluation criteria used by AI.

\subsubsection{Enhanced Selection Process:}

The integration of AI helped streamline the selection process by providing quick, data-driven insights that complemented human expertise.

The methodology ensured a balanced workload among reviewers and maintained thoroughness in the evaluation process.

AI's ability to analyze large datasets quickly and identify patterns provided a valuable second opinion that enhanced the robustness of the selection process.

In cases where AI scores were significantly higher than human scores, such as Teams G and O, AI may have identified potential that human judges overlooked, highlighting the importance of integrating diverse evaluation methods.

The combination of AI and human insights enhances the robustness of the selection process. AI provides a valuable second opinion, highlighting potential areas of interest that might be overlooked by human evaluators.

\subsubsection{Potential for Continuous Improvement:}

The findings suggest areas for improving AI models to better align with human intuition and judgment.

Continuous refinement and training of AI systems based on feedback and discrepancies observed can enhance future evaluations.

Ongoing refinement of AI models is necessary to better align them with human judgment, especially in areas where human intuition plays a critical role.

\subsubsection{Limitations}

Selection Bias: This process led to humans reviewing 36 of the 57 applicants, potentially missing a high-potential startup from the remaining 21 applicants. This is relevant to true life, where investors may use AI tools to bubble up the top startups, but perhaps ignore or not see other high potential startups due to this.

We don’t know the long-term investment payoffs/ROIs of these startup investments/matches. 

Access to Information: The AI did not review the startups’ slide decks, only their spreadsheet of responses from the application form. The human judges had access to the slide decks (presentation PDF or PPT) but not the spreadsheet of responses from the application form.  We could improve this in the future by using an AI that could review the startups’ slide decks, and provide judges with access to the spreadsheets.

\subsection{Finals}

Many people complimented the AI judge inclusion and process.

Startups naturally addressed the AI as “AI Tod” (the AI was made to look and sound like Tod Hynes), and spoke to it, e.g. “Thank you for the great question, AI Tod” and then answered the question

Limitations included the following. First, access to Information: The human judges had the pitch decks, whereas the AI did not (due to limitations of the tool) - and the human judges noted that the AI scores on team composition didn’t align. The tool did not measure timbre of voice since it was using transcripts.

Second, some participants and audience members described the AI as somewhat ‘creepy’

Finally, self-consistency: While the AI scorer scored startup B higher than A, the AI deliberator chose A over B (with the same input)

\subsection{Evaluating the Integration of AI Judges in the ClimaTech Great Global Innovation Challenge}

The ClimaTech Great Global Innovation Challenge aimed to identify and support innovative climate tech startups by incorporating an AI judge into its evaluation process. This analysis delves into the effectiveness, reception, and implications of this novel approach, providing a comprehensive look at the future potential of AI in startup competitions.

\subsubsection{Perception and Awareness}

A key insight from the analysis is the level of awareness and perception of the AI judge among participants. A significant portion were aware that an AI judge was used in the competition, reflecting effective communication by the event organizers. However, the perception of the AI judge's value and fairness was mixed. While some respondents appreciated the AI's ability to pose insightful and tailored questions, others expressed concerns about its decision-making process and potential redundancy with human judges. These mixed perceptions highlight the need for clearer communication regarding the AI's role and decision-making criteria to enhance transparency and trust among participants.

\subsubsection{Fairness and Transparency}

The fairness and transparency of the AI judge's decisions were crucial aspects examined. Participants' ratings revealed a divided opinion, with some rating the AI's fairness highly but expressing lower satisfaction with its transparency. This dichotomy underscores a common challenge in integrating AI into evaluative processes: balancing algorithmic efficiency and human understanding. Participants noted that while the AI provided a novel and impartial perspective, it sometimes failed to consider the context provided by human judges, leading to repetitive or seemingly disconnected questions.

To address this, future implementations could benefit from providing more detailed explanations of how AI decisions are made, perhaps through real-time feedback or post-event debriefs. This approach would not only improve transparency but also help participants understand and trust the AI's contributions. Furthermore, ensuring that the AI system has access to the same information as human judges, such as presentation decks and supplementary materials, would likely enhance its evaluation capabilities and credibility.

\subsubsection{Recommendations for Improvement}

The feedback collected also included several recommendations for enhancing the AI judge's role in future competitions. One recurring suggestion was to improve the AI's integration with human judges. Participants noted that the AI often repeated questions that human judges had already asked, indicating a need for better synchronization between AI and human evaluations. Additionally, there were calls for the AI to have access to the same information as human judges, such as presentation decks and other supplementary materials. This would ensure that the AI's evaluations are as informed and comprehensive as possible, thereby increasing their credibility and usefulness.

Moreover, participants recommended that the AI's role be clearly defined to avoid confusion and overlap with human judges. For instance, the AI could focus on specific areas of evaluation where it excels, such as analyzing large datasets or providing impartial assessments, while human judges handle more nuanced aspects of the evaluation process. This complementary approach would leverage the strengths of both AI and human judges, resulting in a more balanced and effective evaluation process.

The integration of an AI judge in the ClimaTech Great Global Innovation Challenge represents a significant step towards modernizing and enhancing the startup evaluation process. The insights gathered highlight important areas for improvement, particularly in terms of transparency, fairness, and integration with human judges. By addressing these issues, future competitions can leverage AI to provide more robust, objective, and efficient evaluations, ultimately supporting the growth and success of innovative climate tech startups. The ClimaTech initiative, therefore, serves as a valuable case study in the evolving landscape of AI-assisted evaluations, offering lessons that can be applied to a wide range of contexts in the future.

\section{Discussion}

The ClimaTech Great Global Innovation Challenge competition case study demonstrates the practical benefits and challenges of integrating AI into the startup evaluation process. One of the main benefits of the AI judge as used in this process was filtering a potentially large number of applications (in this case 57, but it could have been a much larger number) so the human judges could review a reasonable number. Review and precautions must be taken to minimize the odds that a strong finalist candidate is filtered out before making it to the human review stage. In this case, this risk was mainly addressed by having 36 teams reviewed by humans compared to the final 4 selected. 

AI can identify promising startups and provide a data-driven perspective that complements human expertise. However, discrepancies between human and AI evaluations suggest that ongoing refinement of AI models is necessary to improve their alignment with human judgment. We can also review performance of the various selections over time to see if the human or AI judges perform better at selecting finalists and winners.

This study did not specifically address sex and/or gender dimensions due to its focus on technological and process evaluations. Future research should consider these aspects to enhance generalizability. 

\section{Conclusion}

The integration of AI with human evaluations in the ClimaTech competition provided a robust framework for assessing climate tech startups. This hybrid evaluation model, combining human judgment with AI insights, enhances the accuracy and reliability of startup assessments. Future competitions can refine and expand this model to support more impactful and innovative climate solutions.

This case study illustrates the value of combining human judgment with AI insights in climate tech venture investing. By leveraging both approaches, investors can achieve a more comprehensive evaluation of potential investments, enhancing decision-making processes and ultimately supporting more impactful and innovative climate solutions.

\section*{Declaration of Generative AI and AI-assisted technologies in the writing process}

During the preparation of this work, the authors used OpenAI's GPT-4o and StackAI in order to assist with data analysis and to provide initial insights. After using these tools, the authors reviewed and edited the content as needed and take full responsibility for the content of the publication.

\section*{CRediT Author Statement}

Jennifer Turliuk: Conceptualization; Methodology; Software; Investigation; Writing - Original Draft Preparation.

Alejandro Sevilla: Conceptualization; Methodology; Formal analysis; Investigation; Data Curation; Validation; Visualization; Writing - Original Draft Preparation.

Daniela Gorza: Conceptualization; Writing - Review \& Editing; Validation.

Tod Hynes: Conceptualization; Supervision; Funding Acquisition; Resources; Writing - Review \& Editing.

\section*{Acknowledgements}

This research was supported by Tod Hynes. The funding source had no involvement in the study design, data collection, analysis, interpretation, writing, or decision to submit the article for publication.

%Bibliography
\bibliographystyle{unsrt}  
\bibliography{main}  

\appendix

\section*{Appendices}

\section{Detailed Raw Data}
\label{sec:A}

\subsection{Semi-Finals}

\subsubsection{Compiled Results (AI + Humans)}

The composite scores were derived from a weighted average of human and AI scores.

The table below  (Table \ref{tab:composite_scores}) highlights key data points.

\subsubsection{Humans vs AI}

The comparison between human and AI rankings using Z-scores provided insights into the correlation and differences between these two evaluation methods.

{\fontsize{6pt}{6pt}\selectfont
\begin{longtable}{l *{9}{c}}
\caption{Composite Scores and Ranks}
\label{tab:composite_scores} \\
\toprule
\textbf{Team} & \textbf{Track} & \textbf{Composite Score} & \textbf{Composite Z-Score} & \textbf{Average Human Score} & \textbf{AI Score} & \textbf{Average Human   Z-Score} & \textbf{AI Z-score} & \textbf{Human Rank} & \textbf{AI Rank} \\
\midrule
A             & Early          & 4.6                      & 1.4                        & 4.7                          & 4.5               & 1.5                              & 1                   & 1                   & 4                \\
B             & Early          & 4.3                      & 0.8                        & 4.2                          & 4.5               & 0.7                              & 1                   & 8                   & 4                \\
C             & Early          & 4.3                      & 0.7                        & 4.2                          & 4.5               & 0.6                              & 1                   & 9                   & 4                \\
D             & Early          & 4.2                      & 0.8                        & 4.2                          & 4.2               & 1                                & 0.2                 & 3                   & 12               \\
E             & Late           & 4.2                      & 1.4                        & 3.9                          & 5                 & 1.1                              & 2.3                 & 2                   & 1                \\
F             & Early          & 4.2                      & 0.6                        & 4.3                          & 4                 & 0.9                              & -0.4                & 4                   & 14               \\
\bottomrule
\end{longtable}
}

\subsubsection{Raw Data: }

1. Output from AI (Table \ref{tab:ai_raw_data})

{\fontsize{7.6pt}{7.5pt}\selectfont
\begin{longtable}{l *{7}{c}}
\caption{AI Raw Data and Scores}
\label{tab:ai_raw_data} \\
\toprule
\textbf{Team} & \textbf{Climate   Impact} & \textbf{Solution   Creativity} & \textbf{Idea   Feasibility} & \textbf{Team   Composition} & \textbf{Average   AI Weighted Score} & \textbf{Z-score} & \textbf{Rank} \\
\midrule
A             & 5                         & 5                              & 4                           & 4                           & 4.5                                  & 1.25             & 4             \\
B             & 5                         & 5                              & 4                           & 4                           & 4.5                                  & 1.25             & 4             \\
C             & 5                         & 5                              & 4                           & 4                           & 4.5                                  & 1.25             & 4             \\
D             & 4                         & 4                              & 4                           & 5                           & 4.2                                  & 0.67             & 12            \\
E             & 5                         & 5                              & 5                           & 5                           & 5                                    & 2.23             & 1             \\
F             & 5                         & 4                              & 3                           & 4                           & 4                                    & 0.27             & 14            \\
\bottomrule
\end{longtable}
}

\newpage

2.	Raw data from humans

{\fontsize{6.4pt}{7pt}\selectfont
\begin{longtable}{llllllllll}
\caption{Scoring Rubrics and Criteria}
\label{tab:my-table}\\
\toprule
\textbf{}      & \textbf{}     & \multicolumn{4}{c}{\textbf{Raw}}                                                                          & \multicolumn{4}{c}{\textbf{Processed}}                                                      \\
\midrule
\endfirsthead
\toprule
\textbf{}      & \textbf{}     & \multicolumn{4}{c}{\textbf{Raw}}                                                                          & \multicolumn{4}{c}{\textbf{Processed}}                                                      \\
\midrule
\endhead
\midrule
\multicolumn{10}{r}{\textit{Continued on next page}}\\
\endfoot
\bottomrule
\endlastfoot
\textbf{Judge} & \textbf{Team} & \textbf{Climate Impact} & \textbf{Solution Creativity} & \textbf{Feasibility} & \textbf{Team Composition} & \textbf{Weighted Score} & \textbf{Judge Mean} & \textbf{Judge Std} & \textbf{Judge Z-Score} \\ \midrule
AAA            & O             & 3                       & 4                            & 4                    & 4                         & 3.7                     & 3.5                 & 0.8                & 0.3                    \\
AAA            & W             & 4                       & 4                            & 4                    & 4                         & 4                       & 3.5                 & 0.8                & 0.7                    \\
AAA            & C             & 5                       & 5                            & 4                    & 5                         & 4.7                     & 3.5                 & 0.8                & 1.5                    \\
AAA            & M             & 2                       & 5                            & 4                    & 3                         & 3.4                     & 3.5                 & 0.8                & -0.1                   \\
AAA            & V             & 3                       & 2                            & 2                    & 2                         & 2.3                     & 3.5                 & 0.8                & -1.4                   \\
AAA            & B             & 5                       & 3                            & 3                    & 3                         & 3.6                     & 3.5                 & 0.8                & 0.2                    \\
AAA            & Z             & 2                       & 2                            & 3                    & 3                         & 2.5                     & 3.5                 & 0.8                & -1.2                   \\
\end{longtable}
}

\subsection{Finals}

Raw Data (Table \ref{tab:finals_raw_data})

{\fontsize{8.3pt}{8pt}\selectfont
\begin{longtable}{lllllll}
\caption{Finals Raw Data}
\label{tab:finals_raw_data}\\
\toprule
\textbf{Judge   Name} & \textbf{Team   Name} & \textbf{Climate   Impact} & \textbf{Solution   Creativity} & \textbf{Idea   Feasibility} & \textbf{Team   Composition} & \textbf{Weighted   Score} \\ \midrule
LLL                   & B                    & 5                         & 1                              & 2                           & 3                           & 2.7                       \\
LLL                   & A                    & 5                         & 2                              & 4                           & 4                           & 3.8                       \\
LLL                   & G                    & 3                         & 3                              & 4                           & 1                           & 2.7                       \\
LLL                   & E                    & 5                         & 3                              & 4                           & 5                           & 4.3                       \\
MMM                   & B                    & 5                         & 3                              & 3                           & 3                           & 3.4                       \\
MMM                   & A                    & 5                         & 4                              & 5                           & 5                           & 4.8                       \\
\bottomrule
\end{longtable}
}

\section{Scoring rubrics and criteria}
\label{sec:B}

{\footnotesize
\begin{longtable}{| p{2cm} | p{6.5cm} | p{6.5cm} |}
\hline
\multicolumn{3}{|l|}{\textbf{Judging Criteria}} \\
\hline
\textbf{Category} & \textbf{Description} & \textbf{Scoring (1/low-5/high)} \\
\hline
\multirow{5}{2cm}{\textbf{Potential} \newline \textbf{Climate Impact} \newline \textbf{(20\%)}} & \multirow{5}{6.5cm}{What is the startup’s potential environmental benefit in terms of:
\begin{itemize}[leftmargin=*]
    \item Greenhouse gas (GHG) emissions reduction
    \item Waste reduction
    \item Carbon removal
    \item Lives affected
\end{itemize}
Is the solution needed \textit{today?}
} & 5: High impact. Ex: 100 million people possibly affected; >1 Gigaton of CO\textsubscript{2} eq.; >10\% reduction in water usage; Log 5 Contaminant Removal (99.999\%) \\
\cline{3-3}
& & 3: Medium impact: Ex: 10 million people possibly affected; <1 Gigaton of CO\textsubscript{2} eq.; 5-10\% reduction in water usage; Log 4 to 5 Contaminant Removal (99.99 to 99.999\%) \\
\cline{3-3}
& & 1: Low impact: limited potential towards addressing climate change \\
\hline
\multirow{3}{2cm}{\textbf{Solution} \newline \textbf{Creativity} \newline \textbf{(20\%)}} & \multirow{3}{6.5cm}{
\begin{minipage}{6.5cm}
\vspace{-0.25cm}
\begin{itemize}[leftmargin=*,topsep=0pt]
    \item How innovative and creative is the startup?
    \item To what degree can the idea change the industry?
\end{itemize}
\end{minipage}
} & 5: This is an unseen startup idea with the potential to disrupt a sector/market \\
\cline{3-3}
& & 3: The startup is building upon and improving an existing idea \\
\cline{3-3}
& & 1: The startup is very similar to an existing company/solution \\
\hline
\end{longtable}
}

\section{Additional resources and references}
\label{sec:C}

Climatech GGIC bots: \href{https://docs.google.com/document/d/1u7SVGk83T77yNNMkpbUj9JL8BMZcMUBQH9lSpjGB-tQ/edit\#heading=h.dwgeoebkahei}{https://docs.google.com/document/d/1u7SVGk83T77yNNMkpbUj9JL8BMZcMUBQH9lSpjGB-tQ/edit\#heading=h.dwgeoebkahei} 

Explanation and consent clause: \href{https://docs.google.com/document/d/1OpMgAwGtGd-G-tBnq2j9M1HJnCRYS0f1gKJHHaR6Ct0/edit}{\seqsplit{https://docs.google.com/document/d/1OpMgAwGtGd-G-tBnq2j9M1HJnCRYS0f1gKJHHaR6Ct0/edit}} 

Transcription and tool instructions: \href{https://docs.google.com/document/d/1w-RbCZJJ0YAFrFvyxCrXZrCvtf3TpShOuQFu-1DsW-I/edit}{\seqsplit{https://docs.google.com/document/d/1w-RbCZJJ0YAFrFvyxCrXZrCvtf3TpShOuQFu-1DsW-I/edit}} 

AI prompts: \href{https://docs.google.com/document/d/13OLVqIbtcfN0Ukv4GvbGhAfFD3uCt6kjfGTMr6tSu6g/edit\#heading=h.v1m19k8ueshj}{\seqsplit{https://docs.google.com/document/d/13OLVqIbtcfN0Ukv4GvbGhAfFD3uCt6kjfGTMr6tSu6g/edit\#heading=h.v1m19k8ueshj}} 

\end{document}